\documentclass[reprint,showpacs,preprintnumbers,amsmath,amssymb,prc,floatfix]{revtex4-1}
\usepackage{color}
\usepackage{adjustbox}
\usepackage[mathcal]{eucal}
\usepackage{graphicx}
\usepackage{dcolumn}
\usepackage{bm}
\usepackage{xcolor}
\RequirePackage[colorlinks,citecolor=blue,linkcolor=red,anchorcolor=blue,filecolor=blue,urlcolor=blue]{hyperref}

\begin{document}
\title{Exploring the effect of positive $Q$-value neutron transfer in the Coupled channel calculations using the microscopic nuclear potentials}
\author{N. Jain$^{1}$}
\email{nishujain1003@gmail.com}
\author{M. Bhuyan$^{2}$}
\email{bunuphy@um.edu.my}
\author{P. Mohr$^3$}
\email{mohr@atomki.hu}
\author{Raj Kumar$^1$}
\email{rajkumar@thapar.edu}

\bigskip
\affiliation{$^1$Department of Physics and Materials Science, Thapar Institute of Engineering and Technology, Patiala 147004, India}
\affiliation{$^2$Center for Theoretical and Computational Physics, Department of Physics, Faculty of Science, Universiti Malaya, Kuala Lumpur 50603, Malaysia}
\affiliation{$^3$Institute of Nuclear Research (ATOMKI), H-4001 Debrecen, Hungary}
\date{\today}

\begin{abstract}
We investigated the effect of the degree of freedom of neutron transfer on the cross section of heavy-ion fusion reactions, using the relativistic mean-field formalism within the coupled channel approach (CCFULL). We obtain the microscopic nuclear interaction potential in terms of the density distributions for the targets and projectiles with the NL3$^*$ parameter set and corresponding R3Y nucleon-nucleon potential. The present analysis includes the $^{18}$O-induced reactions, namely, $^{18}$O + $^{58,60,64}${Ni}, $^{18}$O + $^{74}${Ge}, $^{18}$O + $^{148}${Nd}, $^{18}$O + $^{150}${Sm}, and $^{18}$O + $^{182,184,186}$W for which experimental fusion cross-section is available around the Coulomb barrier. It is evident from the results that including vibrational and/or rotational degrees of freedom enhances the fusion cross-section at energies below the barrier. However, fusion hindrance persists in this energy region. To address this, we incorporated the two-neutron $(2n)$ transfer channels in the Coupled Channel calculation. A comparison with the Woods-Saxon potential (WS) shows that the R3Y nucleon-nucleon (NN) potential, with intrinsic degrees of freedom, is superior to it, especially at energies below the barrier. This superiority can be attributed to the observed higher barrier heights and lower cross-section of the WS potential compared to the relativistic R3Y NN potential for the considered reaction systems. Consequently, we employed the relativistic mean-field formalism to estimate fusion characteristics for the unknown $^{18}$O-induced reactions, namely $^{18}$O + $^{62}${Ni}, $^{18}$O + $^{70,72,76}${Ge}, $^{18}$O + $^{144,150}${Nd}, and $^{18}$O + $^{144,148,152,154}${Sm}. Our analysis highlights the significant role of positive $Q$-value neutron transfer in enhancing the sub-barrier fusion cross-section for the $^{18}$O + $^{148}${Nd} reaction with the R3Y NN potential. However, the effect of this transfer channel for the other considered reactions is comparatively less pronounced.
\end{abstract}

\maketitle
\section{Introduction} \label{introduction}
\noindent
The production of exotic nuclei depends on heavy-ion fusion and multi-nucleon transfer reactions, which can be achieved through highly efficient and selective separators, state-of-the-art detectors and sizable primary beam intensities \cite{Comm00,wata15}. These methods are crucial to enhancing our understanding of the nuclear landscape. The fusion of light nuclei is a significant mechanism for energy production in stars, contributing to stellar nucleosynthesis and primordial nucleosynthesis \cite{wall97}. The richness of the comprehensive information regarding the influences of the colliding nuclei's nuclear structure is demonstrated by the sub-barrier fusion in heavy-ion $(A \geq 4)$  collisions. Thus, it prompts physicists to look into the cause of the sub-barrier fusion enhancement that has been reported \cite{Comm00,wata15,wall97}. The combination of an attractive nuclear and repulsive Coulomb potential generates a barrier during collision. Classically, the fusion process can only occur when the incident projectile has more energy than the Coulomb barrier. For systems with relatively low charge values ($Z_1Z_2 \leq 250$), the probability of fusion below the barrier can be accurately characterized using the one-dimensional barrier penetration model (1D-BPM)  \cite{bala98,vaz81,back14,dasg98,wu90,mont17}. However, significant enhancement has been observed beyond the 1D-BPM in the fusion probability below the barrier, especially for heavier systems. These enhancements have been attributed to the interplay of additional degrees of freedom, such as vibrational and rotational excitations, neutron transfer channels, and neck formation \cite{bala98,dass83,hagi12,rajb16,colu19,gaut15,gaut16,saha96,gaut20,rowl92,agui87,sarg12,sarg11} which introduce modifications to the single Coulomb barrier, resulting in a distribution of barriers \cite{bala98,dasg98}. So far, the description of couplings to inelastic excitations in near-barrier fusion has been effectively achieved by applying the Coupled channels (CC) approach. Therefore, the study of near- and sub-barrier heavy-ion fusion provides an excellent platform to explore the mechanisms of coupling and quantum tunnelling \cite{back14,hagi12}. The theoretical models have highlighted the significance of nuclear vibration and rotation in enhancing sub-barrier fusion \cite{hagi12,hagi99,dass87}. However, the understanding of coupling to nucleon transfers remains a challenge, with both microscopic and macroscopic perspectives yet to provide comprehensive insights. 

Interestingly, the effect of transfer reactions on fusion is more pronounced for neutron transfers than for proton transfers, mainly due to the absence of the Coulomb barrier \cite{deb20}. Beckerman \textit{et al.} \cite{beck80} initially discovered a significant influence of neutron transfer with positive $Q$-value on the near-barrier fusion through experimental analysis of fusion excitation functions for $^{58,64}$Ni + $^{58,64}$Ni systems.  Subsequently, this observation gained widespread explanation by attributing the gained kinematic energy of the intermediate states, facilitated by positive $Q$-value neutron transfer (PQNT), as a crucial pathway leading to fusion \cite{lee84,dass94}. Furthermore, various researchers have extensively explored the influence of neutron transfer channels with positive $Q$-values on fusion enhancement \cite{beck80,brog83,brogl83,henn87,stel90,zagr03,jia16,kohl13,rani23}. The effect of PQNT channels was confirmed by experiments of systems $^{32}$Si + $^{100}$Mo \cite{henn91}, $^{40}$Ca +$^{90,96}$Zr \cite{timm98,stef06}, $^{32}$Si + $^{110}$Pd \cite{zhan10}, $^{40}$Ca +$^{124,132}$Sn \cite{kola12}, $^{40}$Ca +$^{70}$Zn \cite{khus19}, etc. Nevertheless, the precise relationship between fusion and neutron transfer remains elusive. Theoretical investigations have revealed alternative mechanisms that can account for the observed experimental results without explicitly incorporating the effect of PQNT. The coupled-channels (CC) method, mentioned earlier, is not the only theoretical framework capable of reproducing fusion excitation functions. Other models, such as the quantum molecular dynamics model and the time-dependent Hartree-Fock method, have also successfully reproduced fusion excitation functions without explicitly considering the intricacies of couplings \cite{wang04,kese12}. Various theoretical investigations \cite{hagi12,hagi99,dass87} sought to identify the cause of the fusion enhancement. 

In their quest to understand the influence of neutron transfer on fusion excitation functions, Jiang \textit{et al.} \cite{jian14} conducted a systematic study. They have found that the fusion excitation function between neutron-deficient projectiles and neutron-rich targets always exhibits a shallower slope than those between pure neutron-rich or neutron-deficient projectile-target systems. It is observed from the recent study \cite{praj23} that the measured sub-barrier fusion cross-section for the $^{28}$Si + $^{158}$Gd reaction is more than that of the $^{30}$Si + $^{156}$Gd reaction due to the significant contribution of positive $Q$-value neutron transfer. Rachkov \textit{et al.} \cite{rach14} highlighted that the PQNT channel substantially influences sub-barrier fusion when the system possesses large positive $Q$-values for neutron transfer and weak coupling to collective states at sub-barrier energies. Furthermore, the systematic investigations carried out by Zhang \textit{et al.} \cite{zhan14} on systems involving PQNT have demonstrated that fusion enhancement occurs as the deformation of the interacting nuclei increases and the mass asymmetry of the system decreases. According to Kohley \textit{et al.} \cite{kohl11}, the influence of transfer channels on the fusion of $Sn$ + $Ni$ and $Te$ + $Ni$ is very weak, with no significant differences observed in the reduced excitation functions. The role of neutron transfer reactions with positive $Q$-values in nuclear fusion is of great importance, primarily because the Coulomb barrier has a limited effect on neutrons. However, several systems, including $^{18}$O + $^{118}$Sn \cite{jaco86}, $^{17}$O + $^{144}$Sm \cite{leig95}, $^{58,64}$Ni + $^{130}$Te \cite{kohl11}, $^{58,64}$Ni + $^{132}$Sn \cite{kohl11}, $^{60,64}$Ni + $^{100}$Mo \cite{stef13}, $^{16,18}$O + $^{76,74}$Ge \cite{lin14}, etc., do not show a significant enhancement in the sub-barrier fusion even with the inclusion of PQNT channel. Furthermore, studies on weakly bound nuclei have revealed that coupling to breakup channels enhances fusion cross sections at sub-barrier energies while reducing them at energies above the barrier \cite{zagr03,dasg04,hagi00,cant09}. However, accurately determining the involvement of weakly bound nucleons in sub-barrier fusion poses a challenge due to the need to consider decay channels and nucleon transfer in complete fusion simultaneously \cite{dasg04,rach13,sarg2012}. 

Several coupled channel models have been developed to compute fusion cross-sections while considering various effects \cite{hagi99,dass87,fern89,dasg92,kuma14,kuma15}. The radial behavior of the nuclear potential shows significant variability, requiring the choice of an appropriate nucleus-nucleus potential to comprehend the complexities of heavy-ion fusion reactions. Among the commonly employed methods is the double-folding approach, which integrates an effective nucleon-nucleon (NN) interaction and nuclear densities to derive the ion-ion optical potential \cite{sing12,sahu14,bhuy14}. Within this framework, several nuclear density models such as the Woods-Saxon ansatz, two-parameter Fermi (2pF), three-parameter Fermi (3pF), Skyrme Hartree Fock (SHF), and the M3Y effective NN potential are utilized to compute the nuclear potential \cite{vrie87,kuma11,jain12,jain14,sick74,raj11,cham02,fold} alongside relevant literature. The M3Y potential, a non-relativistic model, combines the one-pion exchange potential (OPEP) and Yukawa terms to accurately reproduce G-matrix elements in an oscillator basis \cite{satc79}. Conversely, the Woods-Saxon form \cite{hagi99} of the nuclear potential finds frequent use in coupled channel models for understanding heavy-ion reaction mechanisms. Although more intricate than some alternatives, the Woods-Saxon potential is comparatively simpler than fully realistic models, facilitating easier mathematical and computational handling, leading to more straightforward analysis and interpretation of experimental data. Its adjustable parameters can be tuned to fit experimental data, allowing experimentalists to utilize it as a predictive tool for understanding atomic nucleus behavior under different conditions \cite{dalv24}. However, when studying the fusion dynamics of exotic nuclei far from the $\beta$-stable region of the nuclear chart, a microscopic nuclear potential becomes essential \cite{rana22}.

At the microscopic level, several approaches have been employed, such as Skyrme-Hartree-Fock (SHF) \cite{Rein00} and the relativistic mean-field model (RMF) \cite{ring96,bhuy18a,bhuy15,rein89,bhuy20}, to generate nuclear potentials. Recently, the relativistic R3Y effective NN potential has been derived from the self-consistent relativistic mean-field (RMF) approach using specific parameter sets, incorporating meson masses and coupling constants. Additionally, the densities for both interacting nuclei are calculated using the same parameter sets, allowing for their simultaneous utilization in the double-folding approach. This theoretical advancement enables the determination of the nuclear potential through meson interaction while maintaining parameter consistency for both the NN potential and densities. It should be emphasized that RMF models have demonstrated their predictive power in capturing the structural characteristics of finite nuclei, not only in the $\beta$-stable region but also in highly isospin asymmetric regions of the nuclear chart \cite{sing12,sahu14,satc79,ring96,bhuy18a,lahi16,sero86,bhuy15,rein89,bhuy20}. Additionally, the use of the relativistic NN interaction potential, along with the nuclear densities obtained from the RMF formalism, has successfully characterized various nuclear processes, including nuclear fusion, proton radioactivity, and cluster decay \cite{sing12,sahu14,bhuy14}. A compelling method for analyzing low-energy fusion reactions in various systems is to combine the relativistic R3Y potential with the RMF density. 

In our previous studies, we employed the R3Y nucleon-nucleon potentials to investigate the phenomenon of fusion hindrance in specific Ni-based reactions and calculated the cross-sections for synthesizing heavy and superheavy nuclei. We have implemented the R3Y nucleon-nucleon potentials using the $\ell$-summed Wong formula \cite{bhuy20,bhuy18,bhuy22} and references therein. Recently, we have successfully employed the microscopic nuclear potential obtained from the RMF in the Coupled channel calculations to study the effect of the rotational excitation state of the target nuclei on the fusion cross-section \cite{jain2023}. The fusion hindrance is still observed with the inclusion of rotational degrees of freedom. To overcome this fusion hindrance, neutron transfer degrees of freedom are included in the vibrational and/or rotational excitation states of the target. Therefore, in the present work, we intend to include the microscopic nuclear potential obtained from the self-consistent relativistic mean-field approach in the channel coupling approach and then investigate the possible effect of neutron transfer channels on the fusion dynamics concerning those obtained from traditional Woods-Saxon potential. A thorough understanding of PQNT on sub-barrier fusion remains incomplete, despite the use of many theoretical papers in the field of transfer and fusion. Experimental evidence is insufficient to fully understand this concept. In the present study, we have started with the $^{18}$O-based reactions for which experimental data are available, namely, $^{18}$O + $^{64}${Ni}, $^{18}$O + $^{74}${Ge}, $^{18}$O + $^{148}${Nd}, $^{18}$O + $^{150}${Sm}. In the specific case of $^{18}$O induced reactions with $2n$ transfer, the behaviour of target-like nuclei undergoes a noticeable change in deformation \cite{rach14}. Rachkov \textit{et al.'s} investigations highlight the crucial role of neutron rearrangement in enhancing fusion, particularly when colliding nuclei possess magic proton or neutron numbers. Furthermore, we extend this work to reactions for which experimental data are not available, namely, $^{18}$O + $^{62}${Ni}, $^{18}$O + $^{70,72,76}${Ge}, $^{18}$O + $^{144,150}${Nd}, $^{18}$O + $^{144,148,152,154}${Sm}. The sub-barrier fusion cross-section for these reactions is examined by applying microscopic R3Y NN potentials, employing the NL3$^*$ parameter set, and subsequently compared with the results obtained using a static Woods-Saxon potential. The rarity of $^{18}$O-induced reactions in heavy-ion fusion makes these reactions suitable for investigating the effect of neutron transfer on heavy-ion fusion enhancement in energy regions below the barrier. Moreover, a comparative analysis is performed between the theoretical results and experimental data for the considered nuclear reactions.\\ 
The structure of the present work is as follows: The methodology is covered in Sec. \ref{theory}. In Sec. \ref{results}, coupled channel calculations are described, and their results are contrasted with experimental data. The summary and conclusion are presented in Sec. \ref{result}.
\section{Theoretical Formalism} 
\label{theory} 
\noindent
The present work employs the channel coupling (CC) approach based code CCFULL \cite{hagi99}, a multidimensional barrier penetration model that offers valuable insights into the fusion dynamics at energies near the Coulomb barrier. This program determines the mean angular momenta of the compound nucleus and the fusion cross sections under the effect of couplings between the relative motion along with additional nuclear collective motions. The coupled channels equations, which encompass the relevant channels, are typically solved numerically to address the effects of the coupling between relative motion and intrinsic degrees of freedom on fusion \cite{hagi12,hagi99,Esbe87,rumi99}. The CC equation is given as, 
\begin{eqnarray}
\Bigg[\frac{-\hbar^2}{2\mu} \frac{d^2 }{dr^2} &+& \frac{J(J+1)\hbar^2}{2\mu r^2}+\frac{Z_P Z_T e^2}{r}+V_{N}+\epsilon_n -E_{c.m.}\Bigg] \nonumber \\
&& \times \psi_n (r) + \sum_m V_{nm} (r)\psi_m (r) =0. 
\label{cc1}   
\end{eqnarray}
Here, each symbol has its usual meaning. More details about CC calculations can be found in Ref. \cite{hagi99}. All the Coupled channel equations which have non-linear coupling are significant in studying the heavy-ion fusion reactions, mainly at sub-barrier energies. The fusion cross-sections are calculated as demonstrated below to capture the effects of all significant intrinsic channels. The fusion cross-sections are computed by considering the influence of all prominent intrinsic channels, i.e.,
\begin{eqnarray}
\sum_{J}\sigma_J (E)=\sigma_{fus} (E)=\frac{\pi}{k_{0}^{2}}\sum_{J}(2J+1)P_J (E).
\label{fusion}
\end{eqnarray}
To simplify the calculation, the iso-centrifugal approximation is employed \cite{hagi12,hagi99,muha08}, and the total angular momentum {\it J} is substituted with $\ell_{max}$ in each channel. The $\ell_{max}$ is the total angular momentum that contributes to the fusion cross-section. In the present work, the $\ell_{max}$ values are obtained from the sharp cut-off model \cite{beck81} for the above energies and are extrapolated for energies below the Coulomb barrier.\\ \\
The crucial component of the coupled channels approach is the nucleus-nucleus interaction potential, which was taken as a Woods-Saxon form for the known region of the nuclear chart and is given as:
\begin{equation}
    V_{N}= \frac{-V_0}{1+\exp\Big[\big(r-R_0\big)/a_0\Big]},  R_0=r_0(A_{P}^{1/3}+A_{T}^{1/3}),
    \\ \label{ws}
\end{equation}
where $V_0$, $r_0$, and $a_0$ are the nuclear potential parameters. Parallel to the traditional Woods-Saxon (WS) potential, here we employ the nucleus-nucleus potential directly constructed from relativistic R3Y NN potential \cite{sing12,sahu14,bhuy14} using the densities from the RMF approach for the NL3$^*$ parameter set for the investigation of fusion dynamics.
The newly developed R3Y NN potential can be obtained from the static solution of the field equations for mesons \cite{sing12,bhuy14,bhuy18}, and can be written as:
\begin{eqnarray}
V_{\mbox{eff}}^{R3Y}(r)&=&\frac{g^2_\omega}{4\pi}\frac{e^{-m_\omega r}}{r}+\frac{g^2_\rho}{4\pi}\frac{e^{-m_\rho r}}{r} -\frac{g^2_\sigma}{4\pi}\frac{e^{-m_\sigma r}}{r} \nonumber \\
&+&\frac{g^2_2}{4\pi}re^{-2m_\sigma r} +\frac{g^2_3}{4\pi}\frac{e^{-3m_\sigma r}}{r}+J_{00}(E)\delta(r),
\label{r3y}
\end{eqnarray}
Here, the parameters $g_\sigma$, $g_\omega$, $g_\rho$ represent the respective coupling constants of the mesons having masses $m_\sigma$, $m_\omega$ and  $m_\rho$, respectively. The coupling constants $g_2$, and $g_3$ correspond to the non-linear terms of the self-interacting $\sigma$ field. The $J_{00}$ is the one-pion exchange potential (OPEP). More details about R3Y NN potential can be found in Ref. \cite{satc79}. Here we have used a revisited version of the widely used NL3 force \cite{lala97}, so-called NL3$^*$ parameter set \cite{lala09}. The interaction potential between the projectile and target nuclei, considering their respective calculated nuclear densities $\rho_p$ and $\rho_t$ with the RMF approach for NL3$^*$ parameter, can be determined using 
\begin{eqnarray}
V_{n}(\vec{R})&=&\int\rho_{p}(\vec{r}_p)\rho_{t}(\vec{r}_t)V_{eff}  \left( |\vec{r}_p-\vec{r}_t +\vec{R}| {\equiv}r \right) \nonumber \\ 
&& d^{3}r_pd^{3}r_t,
\label{DF}
\end{eqnarray}
the double-folding approach \cite{satc79} with the relativistic R3Y interaction potentials proposed in Refs. \cite{sing12,sahu14,bhuy18}. Additionally, single nucleon exchange effects can be included through a zero-range pseudo-potential.
The total interaction potential between the projectile and target nuclei can be obtained by combining the Coulomb potential with the nuclear interaction potential $V_n(R)$ obtained from Eq. (\ref{DF}),
which is the main ingredient in the Coupled channels approach.\\ \\
The nuclear coupling Hamiltonian can be generated by changing the target radius in the nuclear potential to a dynamical operator,
\begin{equation}
   R_0 \rightarrow R_0 + \hat{O}= R_0 + \beta_2R_TY_{20}+\beta_4R_TY_{40},
\end{equation}
where $R_T$ is parameterised as $r_{coup}A^{1/3}$ and $\beta_2$, $\beta_4$ are the quadrupole and hexadecapole deformation parameters of the deformed target nucleus, respectively. Thus, the nuclear coupling Hamiltonian is given by
\begin{equation}
    V_{N}(r,\hat{O})= \frac{-V_0}{1+\exp\Big[\big(r_0-R_0-\hat{O}\big)/a_0\Big]},  \\ 
    \label{ws1}
\end{equation}
The ground rotational band of the target requires matrix elements of this coupling Hamiltonian between the $|n\rangle=|I0\rangle$ and $|m\rangle=|I'0\rangle$ states. These can be readily achieved through matrix algebra \cite{Kerm93}. In the context of algebraic analysis, the initial step involves the identification of the eigenvalues and eigenvectors associated with the operator $\hat{O}$, which follows the given conditions
\begin{equation}
    \hat{O}\mid \alpha > = \lambda_\alpha\mid \alpha > 
\end{equation}
The CCFULL program diagonalize the matrix $\hat{O}$, whose elements are given by
\begin{eqnarray}
\begin{aligned}
    \hat{O}_{II'}= \sqrt{\frac{5(2I+1)(2I'+1)}{4\pi}}\beta_2R_T
\begin{pmatrix}
I & 2 & I'\\
0 & 0 & 0 
\end{pmatrix}^2 \\
+ \sqrt{\frac{9(2I+1)(2I'+1)}{4\pi}}\beta_4R_T
\begin{pmatrix}
I & 4 & I'\\
0 & 0 & 0 
\end{pmatrix}^2.  
\end{aligned}
\end{eqnarray}
The nuclear coupling matrix elements are then evaluated as
\begin{eqnarray}
V_{nm}^{(N)}&=& <I0\mid V_N(r,\hat{O})\mid I^{'}0> - V_{N}^{(0)}(r)\delta_{n,m}, \nonumber \\
&&  = \sum_\alpha <I0\mid \alpha><\alpha\mid I^{'}0> V_N(r,\lambda_\alpha)-\nonumber \\
&& V_{N}^{(0)}(r)\delta_{n,m}.   
\label{am1}
\end{eqnarray}
The final term in this equation is included to prevent the diagonal component from being counted twice. The program CCFULL incorporates the Coulomb interaction of the deformed target, considering terms up to second order in $\beta_2$ and first order in $\beta_4$. In contrast to nuclear couplings, previous studies have demonstrated that the higher-order couplings associated with the Coulomb interaction have a relatively insignificant effect. The matrix elements are subsequently provided by:
\begin{eqnarray}
    \begin{split}
        V_{nm}^{(C)}=\frac{3Z_PZ_TR^2_T}{5r^3}\sqrt{\frac{5(2I+1)(2I'+1)}{4\pi}}\times \\
        \Bigg( \beta_{2}+\frac{2}{7}\beta^2_2\sqrt{\frac{5}{\pi}}\Bigg)\begin{pmatrix}
I & 2 & I'\\
0 & 0 & 0 
\end{pmatrix}^2 \\+\frac{3Z_PZ_TR^4_T}{9r^5}\sqrt{\frac{9(2I+1)(2I'+1)}{4\pi}}\times\\
\Bigg( \beta_{4}+\frac{9}{7}\beta^2_2\Bigg)\begin{pmatrix}
I & 4 & I'\\
0 & 0 & 0 
\end{pmatrix},
    \end{split}
\end{eqnarray}
for rotational coupling. The total coupling matrix element is obtained by adding the values of $V_{nm}^{(N)}$ and $V_{nm}^{(C)}$.\\ \\
For vibrational coupling, the operator $\hat{O}$ in the
nuclear coupling Hamiltonian is given by
\begin{equation}
    \hat{O}=\frac{\beta_\lambda}{\sqrt{4\pi}}R_T{(a^{\dagger}_{\lambda0} +a_{\lambda0})},
\end{equation}
In the given expression, $\lambda$ represents the multipolarity of the vibrational mode, while ${a^{\dagger}_{\lambda0}}$ and $({a_{\lambda0}})$ denote the creation and annihilation operators of the phonon, respectively. The matrix element of the operator between the n-phonon state $|n\rangle$ and the m-phonon state $|m >$ can be expressed as follows:
\begin{equation}
  \hat{O}_{nm}= \frac{\beta_\lambda}{\sqrt{4\pi}}R_T(\sqrt{m}\delta_{n,m-1} + \sqrt{n}\delta_{n,m+1}.
\end{equation}\\
The subsequent steps involved in determining the nuclear coupling matrix element are identical to those used in the rotational coupling. The operator $\hat{o}$ is diagonalized within a physical space, and afterwards, the nuclear coupling matrix elements are obtained using eq. (\ref{am1}).\\
The CCFULL program employs the linear coupling approximation to simulate the Coulomb coupling of the vibrational degree of freedom. The Coulomb coupling matrix elements are therefore observed:\\
\begin{eqnarray}
    V_{nm}^{(C)}(r) &=&\frac{\beta_\lambda}{\sqrt{4\pi}}\frac{3}{2\lambda+1}Z_PZ_Te^2\frac{R_T^\lambda}{r^{\lambda+1}}\nonumber \\
&& {(\sqrt{m}\delta_{n,m-1} + \sqrt{n}\delta_{n,m+1}).}
\end{eqnarray}
Again, the total coupling matrix element is determined by adding the values of $V_{nm}^{(N)}$ and $V_{nm}^{(C)}$.\\
The CCFULL program utilizes a macroscopic coupling form factor to account for the pair-transfer coupling between the ground states of nuclei, whereby the nuclear coupling Hamiltonian for vibrational and rotational coupling can be generated by replacing the target radius with the dynamical operator in the nuclear potential \cite{dass86}.
\begin{equation}
    F_{trans}(r) = F_{tr} \frac{dV_N}{dr}
    \label{ftr}
\end{equation}
Here $F_{tr}$ is the coupling strength.\\
The above-discussed approach is employed to study the fusion dynamics of $^{18}$O-based reactions and is discussed in the following Section \ref{results}.
\begin{table*} 
\caption{\label{table1} The Woods-Saxon (WS) parameters ($V_0$, $r_0$, $\&$ $a_0$), quadrupole ($\beta_2$) and hexadecapole ($\beta_4$) deformation calculated using RMF  (with NL3$^*$ parameter set), $1n$, and $2n$ stripping $Q$-values are listed. The excitation energy corresponds to quadrupole deformation of the target nuclei is taken from Ref. \cite{rama01}.}
\centering
\renewcommand{\tabcolsep}{0.1cm}
\renewcommand{\arraystretch}{1.5}
\begin{tabular}{|c|ccc|c|cc|cc|cc|c|}
\hline \hline 
System & \multicolumn{3}{c|}{Woods-Saxon Potential} 
 & \multicolumn{1}{c|} {Excitation energy} & \multicolumn{4}{c|}{Deformation} & \multicolumn{2}{c|}{$Q$-value} & Expt.\\
 \cline{2-5}\cline{6-12}
& $V_0$ & $r_0$ & $a_0$ &  \multicolumn{1}{c|} {$E_2^+$} &\multicolumn{2}{c|}{Before $2n$ transfer} & \multicolumn{2}{c|}{after $2n$ transfer} &\multicolumn{1}{c|}{$1n$} &\multicolumn{1}{c|}{$2n$} & \\
& (MeV) & (fm) & (fm) & \multicolumn{1}{c|}{(MeV)} & $\beta_2$ & \multicolumn{1}{c|}{$\beta_4$} &$\beta_2$ & $\beta_4$ &\multicolumn{1}{c|}{ MeV}&\multicolumn{1}{c|}{ MeV} &\\
\hline
$^{18}$O+$^{58}$Ni  &55.35& 1.17 & 0.52 &1.45 & 0.001& 0.001 &0.088& 0.007 &0.95 &8.20 & \cite{silv97}\\
$^{18}$O+$^{60}$Ni  & 55.30& 1.17& 0.72 & 1.33 & 0.088& 0.007  &0.101& -0.008&-0.23& 6.23&\cite{silv97}\\ 
$^{18}$O+$^{62}$Ni & 55.27& 1.17& 0.64 & 1.17  & 0.101& -0.008 & -0.091&-0.002& -1.21&4.31&-\\
$^{18}$O+$^{64}$Ni & 55.25& 1.17 & 0.64 & 1.35 & -0.091& -0.002 &0.0001&0.000 &-1.95&2.86&\cite{silv97}\\
$^{18}$O+$^{70}$Ge & 56.52  & 1.17& 0.65 & 1.04 & 0.212& -0.015 & 0.214& -0.023&-0.63&5.98 &-\\
$^{18}$O+$^{72}$Ge  & 56.49& 1.17& 0.64 & 0.83 & 0.214& -0.023 & 0.213&-0.021&-1.26&4.79 &-\\
$^{18}$O+$^{74}$Ge  & 56.46  & 1.17 & 0.64 & 0.60 & 0.213& -0.021 & 0.163&0.021&-1.54&3.75 &\cite{jia12}\\
$^{18}$O+$^{76}$Ge  & 56.44  & 1.17& 0.64 & 0.56 & 0.163&0.021 & 0.178& 0.002&-1.97&2.60 &-\\
$^{18}$O+$^{144}$Nd  &61.95& 1.18 & 0.66 &0.70 & 0.005& 0.003 & 0.063&0.018&-2.29&1.13 & -\\
$^{18}$O+$^{148}$Nd  &61.89& 1.18& 0.66 & 0.30 & 0.164& 0.049& 0.226&0.079&-3.01&0.225 & \cite{brod75}\\ 
$^{18}$O+$^{150}$Nd & 61.87& 1.18& 0.66 & 0.13 & 0.226& 0.079 &0.283&0.109 &-2.71&0.42 &-\\
$^{18}$O+$^{144}$Sm & 62.31& 1.18 & 0.66 & 1.66 & 0.002& 0.004 & 0.017&0.003&-1.29&2.99 &-\\
$^{18}$O+$^{148}$Sm &62.20  & 1.18& 0.66 & 0.55 & 0.113& 0.049 & 0.181&0.061&-2.18&1.67 &-\\
$^{18}$O+$^{150}$Sm  & 62.21& 1.18& 0.66 & 0.33 & 0.181& 0.061 &0.237 &0.083&-2.45&1.66 &\cite{char86}\\
$^{18}$O+$^{152}$Sm  & 62.18  & 1.18 & 0.66 & 0.12 & 0.237& 0.083 & 0.267&0.098&-2.18&1.65 &-\\
$^{18}$O+$^{154}$Sm  & 62.16  & 1.18& 0.66 & 0.08 & 0.267& 0.098 & 0.293&0.101&-2.24&0.86 &-\\
$^{18}$O+$^{182}$W  & 98.76& 1.15& 0.73 & 0.10 & 0.265& -0.075 & 0.256&-0.087&-1.86&1.41 & \cite{jish22}\\
$^{18}$O+$^{184}$W  & 98.76& 1.15& 0.73 & 0.11 & 0.256& -0.087 &0.243&-0.097 &-2.29&0.76 &\cite{jish22}\\
$^{18}$O+$^{186}$W  & 98.76& 1.15& 0.73 & 0.12 & 0.243& -0.097 & 0.233&-0.078&-2.58&0.11 &\cite{jish22}\\
\hline \hline 
\end{tabular}
\end{table*}
\begin{figure*}
\begin{center}
\includegraphics[width=170mm,height=190mm,scale=1.5]{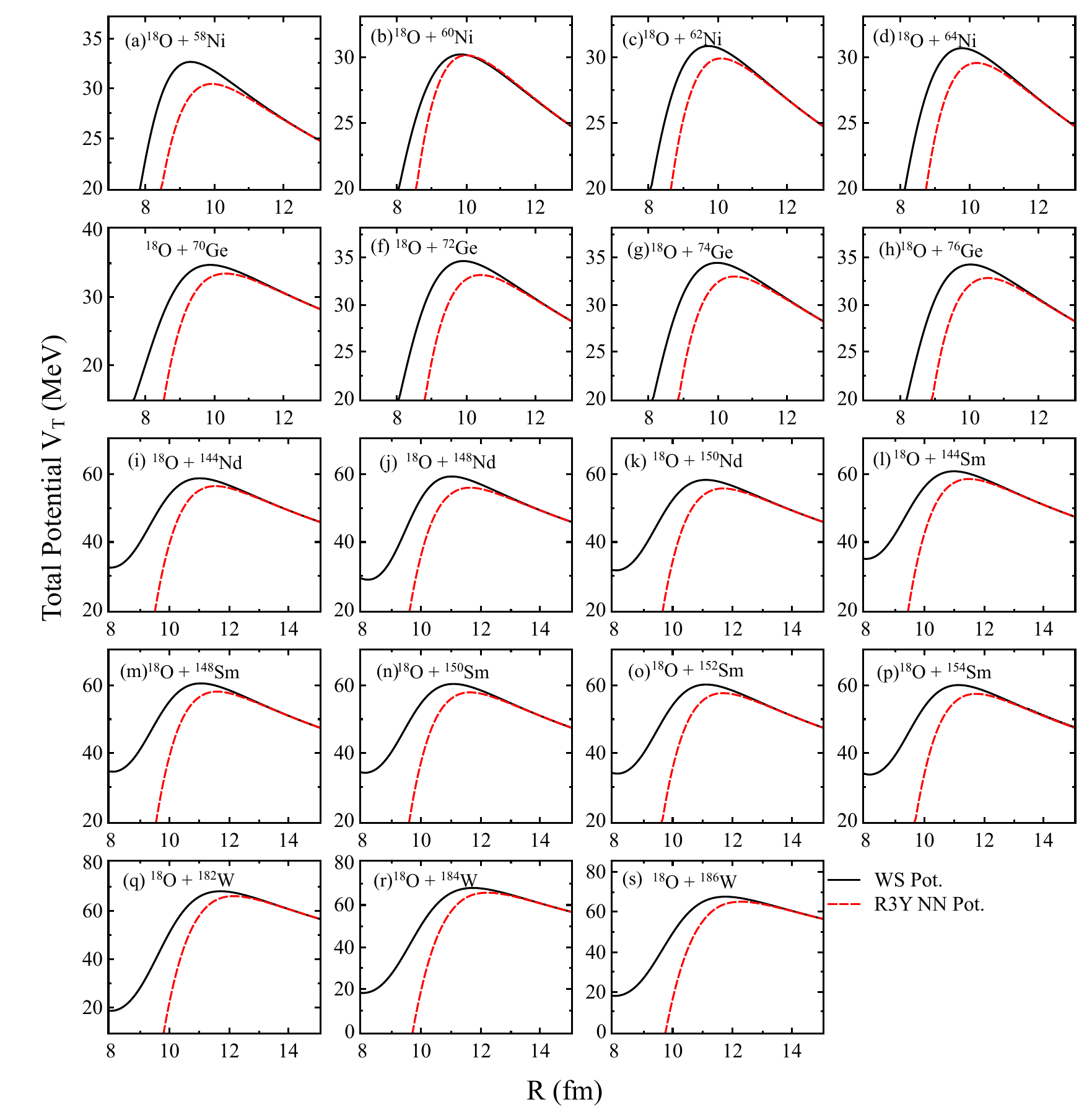}
\vspace{-0.3cm}
\caption{\label{fig. 1} (Color online) Comparison of the radial dependence of the total interaction potential $V_T$ (MeV) at $\ell$ = 0$\hbar$ for the considered reaction using two different potential models - the R3Y NN interaction potential (red dashed line) and the WS potential (solid black line).}
\end{center}
\end{figure*}

\section{Results and Discussions}
\label{results} \noindent
The Coupled channel calculations adequately consider many degrees of freedom to explain the fusion cross-section quite well at lower excitation energies. The following procedure is used in the present study to perform the Coupled channel calculations with microscopic nuclear potential:
\begin{enumerate}
    \item The relativistic R3Y NN potential and the densities from the relativistic mean-field approach are used to obtain the relativistic nuclear interaction potential.
  
    \item The relativistic mean field formalism is used to obtain the 
     structural bulk properties, such as binding energy, nuclear radii, density distribution, quadrupole ($\beta_2$), and hexadecapole ($\beta_4$) deformation, which provides a picture of nuclei involved in the reaction. 
    
    \item The fusion cross-section in coupled channels calculations is determined by incorporating the density and the shape degree of freedom, specifically the quadrupole ($\beta_2$), and hexadecapole ($\beta_4$), into the nuclear potential.
    
    \item The fusion enhancement at below-barrier energies is examined by taking into account the vibrational and/or rotational degrees of freedom up to $2^+$ states.
    
    \item The effect of the positive $Q$-value neutron transfer on the fusion cross-section is explored along with the microscopic nuclear potential at below-barrier energies for the considered choices of reactions.
\end{enumerate}
The motivation for selecting the external nuclear potential derived from RMF formalism is its ability to provide crucial insights into bulk nuclear properties, including binding energy, charge distributions, and single particle energy levels. Further information on the suitability and effectiveness of the RMF model with different parameterizations can be found in Refs. \cite{sing12,bhuy14,bhuy18} and reference therein. The microscopic relativistic R3Y NN interaction potential with NL3$^*$ parameter is used to calculate the fusion cross-section for $^{18}$O-induced reactions, namely, $^{18}$O + $^{58,60,62,64}${Ni}, $^{18}$O + $^{70,72,74,76}${Ge}, $^{18}$O + $^{144,148,150}${Nd}, $^{18}$O + $^{144,148,150,152,154}${Sm}, $^{18}$O + $^{182,184,186}$W by using the quantum mechanical channel-coupling code CCFULL. The target nuclei considered in the reactions mentioned above, i.e., $^{58,60,62,64}${Ni}, $^{70,72,74,76}${Ge}, $^{144}${Nd}, and $^{144,148}${Sm}, are vibrational in nature. On the other hand, $^{148,150}${Nd}, $^{150,152,154}${Sm}, $^{182,184,186}$W target nuclei follow the rotational behavior. For example, in the case of $^{148,150}${Nd}, $^{150,152,154}${Sm}, $^{182,184,186}$W nuclei, the energies of rotational level for I = $2^+$, $4^+$, $6^+$ states are almost comparable to the experimental data \cite{kran86}. However, the main energy band diagram for $^{58,60,62,64}${Ni}, $^{70,72,74,76}${Ge}, $^{144}${Nd}, and $^{144,148}${Sm} nuclei is far from $I(I+1)^*E_2T/6$ expression (for rotational energy). The fusion barrier characteristics, i.e., frequency, height, and barrier position, are obtained using the microscopic R3Y NN interaction potential \cite{bhuy20,shil21}. In addition, the fusion cross-sections obtained with microscopic nuclear potential are further compared with the Woods-Saxon (WS) potential as well.

\subsection{Nuclear Interaction Potential}
The accurate characterization of fusion properties within a system critically depends upon the interaction potential between the two interacting nuclei. The potential parameters of the WS potential are chosen in such a way as to obtain the same fusion cross-section as that of experimental data at above-barrier energies for 1D-BPM. Nonetheless, for the other theoretical reactions being studied, the potential parameters are taken directly from Ref. \cite{fold}. The Aky$\ddot{u}$z-Winther parameterization of WS potential, the excitation energy of the target nuclei, and the corresponding deformations calculated using RMF with NL3$^*$ parameter set \cite{bhuy18} are tabulated in Table \ref{table1}. The total interaction potential is calculated for each considered reaction, as shown in Fig. \ref{fig. 1}. The solid black and red dashed line represents the total interaction potential corresponding to WS, and R3Y NN potential, respectively. On comparing the microscopic nuclear potential with the WS potential, it is observed that the folding potentials seem to be superior to the widely used WS potentials; but maybe, using
different WS parameters might lead to excitation functions close to the folding approach. This is attributed to the more attractive nature of the R3Y NN potential, which considers the meson masses and coupling constants more effectively. As the calculated barrier height for a system increases, its cross section decreases correspondingly. 

\subsection{Fusion Cross-Section}
\subsubsection{Single-Barrier Penetration Model}
The Coupled channel calculations are done to further investigate the impact of the potentials mentioned above on the fusion cross-section. Initially, we start with the nuclear reactions for which experimental fusion cross-section is available in the literature, i.e., $^{18}$O + $^{58,60,64}$Ni, $^{74}$Ge, $^{148}$Nd, $^{150}$Sm, $^{182,184,186}$W \cite{silv97,jia12,brod75,char86,jish22}. The 1D-BPM calculations are carried out without implementing the intrinsic nuclear excitations to reproduce the experimental data at the above barrier energies. The black (dotted-dashed) and blue (dotted-dashed) lines represent the data for 1D-BPM corresponding to WS and R3Y NN interaction potential, as shown in Figs. \ref{fig. 2}(a), \ref{fig. 2}(b), \ref{fig. 2}(d), \ref{fig. 2}(g), and \ref{fig. 3}(c), \ref{fig. 3}(g), \ref{fig. 3}(i), \ref{fig. 3}(j), \ref{fig. 3}(k) for $^{18}$O + $^{58,60,64}$Ni, $^{74}$Ge, $^{148}$Nd, $^{150}$Sm, $^{182,184,186}$W reactions, respectively. The solid black circles in the Figs. \ref{fig. 2}, \ref{fig. 3} represent the experimental fusion cross-section. The sharp cut-off model, as described in Ref. \cite{kuma09}, is used to determine the maximum orbital angular momentum ($\ell_{max}$) values for reaction systems that have experimental data in the above-barrier region \cite{beck81}, which are then extrapolated to sub-barrier energies. The 1D-BPM underestimates the experimental fusion cross-section, mainly at sub-barrier energies. But reproduces the data well at above barrier energies. It is observed from Figs. \ref{fig. 2}, \ref{fig. 3} that the fusion cross-section obtained using R3Y NN potential gives better match with the experimental data compared to the WS potential, mainly at above barrier energies. Therefore, we intend to go one step ahead for the reaction systems for which experimental data is unavailable. Since the sharp cut-off model is not applicable for all the considered systems because the experimental data is not available for them, i.e., $^{18}$O + $^{62}${Ni}, $^{18}$O + $^{70,72,76}${Ge}, $^{18}$O + $^{144,150}${Nd}, $^{18}$O + $^{144,148,152,154}${Sm}. We have used the polynomial between $E_{c.m.}$/$V_b$ and $\ell_{max}$ value generated using $^{18}$O + $^{64}${Ni}, $^{18}$O + $^{74}${Ge}, $^{18}$O + $^{148}${Nd}, $^{18}$O + $^{150}${Sm} (whose experimental data are available) for the corresponding systems, respectively. The calculated fusion cross-section using R3Y NN interaction potential is much higher than the WS potential as shown in Figs. \ref{fig. 2}(c), \ref{fig. 2}(e), \ref{fig. 2}(f), \ref{fig. 2}(h) and \ref{fig. 3}(a), \ref{fig. 3}(c), \ref{fig. 3}(d), \ref{fig. 3}(e), \ref{fig. 3}(g), \ref{fig. 3}(h).

\subsubsection{Effect of vibration and/or rotation of target nuclei on the Fusion Cross-Section}
\begin{figure*}
\begin{center}
\includegraphics[width=170mm,height=190mm,scale=1.5]{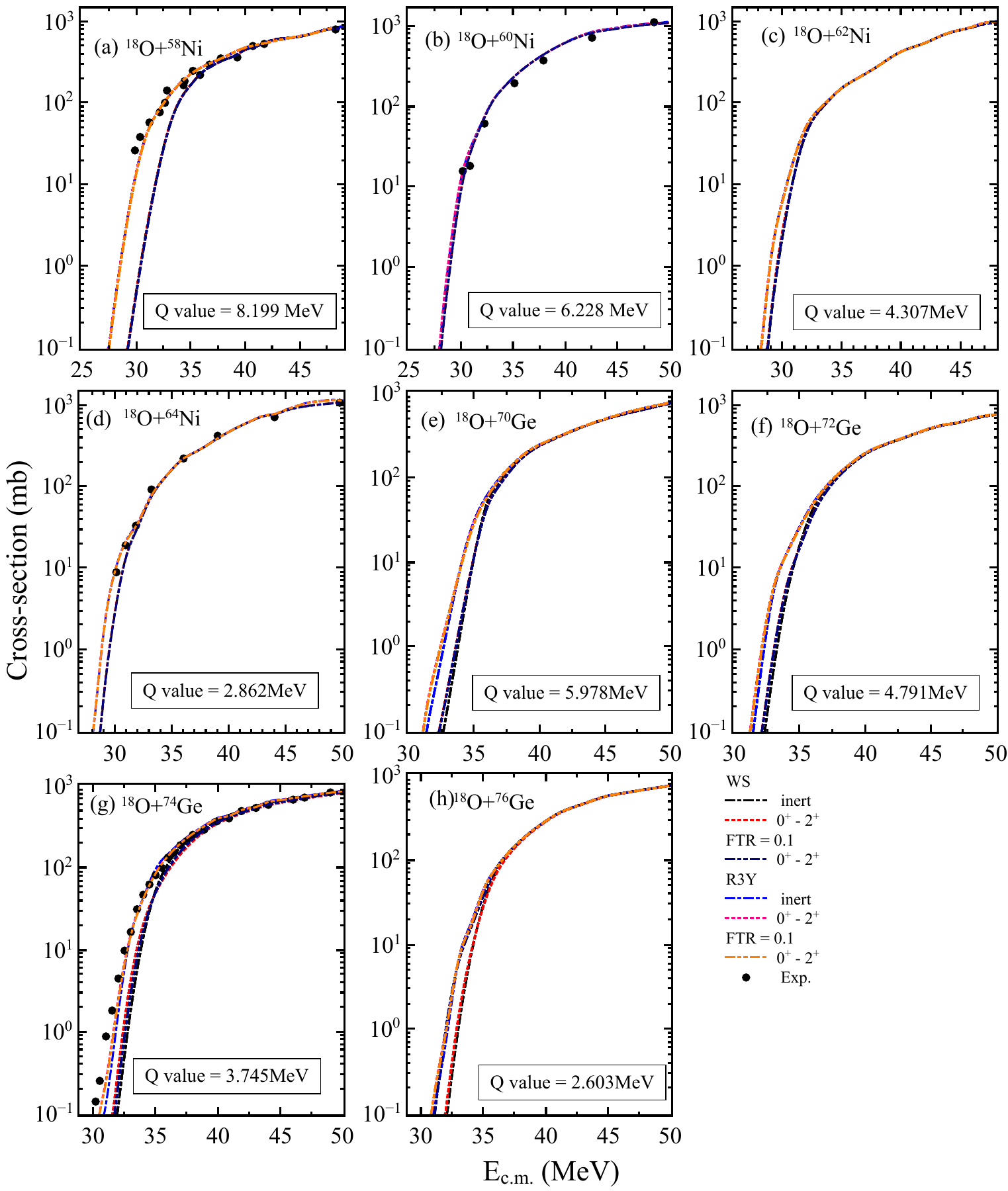}
\vspace{-0.3cm}
\caption{\label{fig. 2} (Color online) The figure shows the fusion excitation functions for the $^{18}$O + $^{58,60,62,64}$Ni and $^{18}$O + $^{70,72,74,76}$Ge reactions obtained using both Woods-Saxon and R3Y NN potential with a sharp cut-off model. The 1D-BPM model Coupled channel calculations are represented by the dashed dotted black and blue lines corresponding to the WS and R3Y NN interaction potentials, respectively. The dotted red (WS) and pink (R3Y NN) lines show the fusion cross-sections attained after including vibrational or rotational states of the target nuclei. The dotted dashed green (WS) and purple (R3Y NN) lines represent the fusion cross-section obtained after including $2n$-neutron transfer with a coupling strength of 0.1. The theoretical data is compared with the experimental data \cite{silv97,jia12}. See text for more details.}
\end{center}
\end{figure*}

\begin{figure*}
\begin{center}
\includegraphics[width=170mm,height=240mm,scale=1.5]{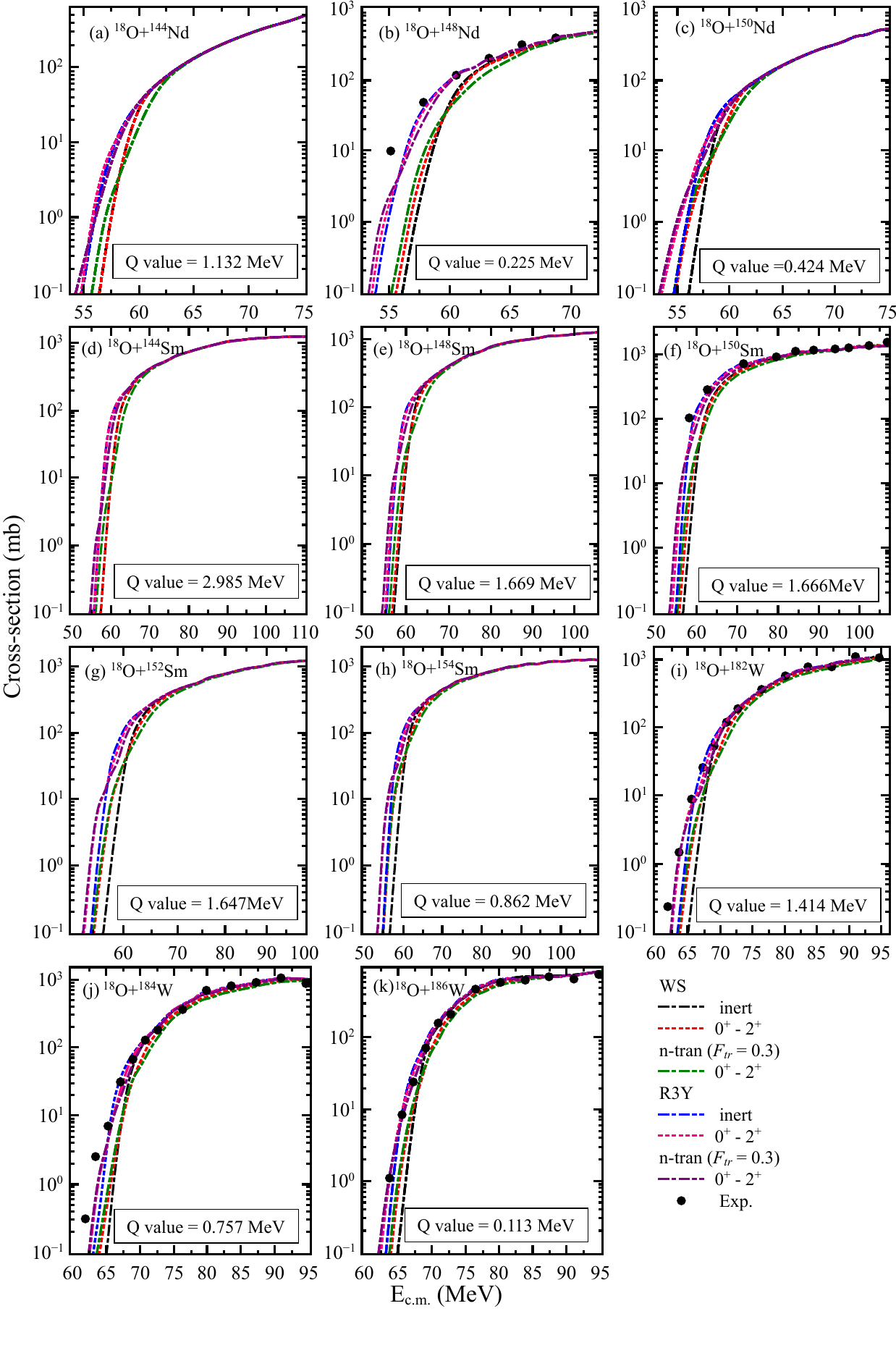}
\vspace{-0.8cm}
\caption{\label{fig. 3} (Color online) Same as figure \ref{fig. 2}, but for $^{18}$O + $^{144,148,150}${Nd}, $^{18}$O + $^{144,148,150,152,154}${Sm}, $^{18}$O + $^{182,184,186}$W reactions. The results are further contrasted with the experimental results as well \cite{brod75,char86,jish22}. See text for more details.}
\end{center}
\end{figure*}

The coupled channel (CC) calculations includes the intrinsic properties of the colliding nuclei i.e., vibrations, deformations, neutron transfer channel without any change in the potential parameters. All known states with strong E2 or E3 transition strength to the ground state were considered while considering the inelastic channels. The CC calculations include the vibrational and/or rotational degrees of freedom to reduce the fusion hindrance at energies below the Coulomb barrier. The inclusion of $\beta_4$ values with $\beta_2$ further enhances the fusion cross-section. Our recent work \cite{jain23} also shows that hexadecapole deformation ($\beta_4$) with different magnitudes significantly influences the fusion cross-sections. Initially, we start with the target nuclei having vibrational $2^+$ state as its first excitation state for $^{58,60,62,64}$Ni, $^{70,72,74,76}$Ge, $^{144}$Nd, and $^{144,148}$Sm nuclei to perform the CC calculations. The dotted red and magenta lines in Figs. \ref{fig. 2}, \ref{fig. 3} represent the fusion cross-section after including the $1^{st}$ excited state of the target nuclei corresponding to the WS and R3Y NN interaction potential. There is no significant enhancement in the fusion cross-section for $^{18}$O + $^{58,60,62,64}$Ni, $^{18}$O + $^{76}$Ge, $^{18}$O + $^{144}$Nd, and $^{18}$O +  $^{144,148}$Sm reactions w.r.t. the 1-D BPM as the values of $\beta_2$ are quite small (between -0.091 to 0.163), as mentioned in Table \ref{table1}. However, the $\beta_2$ value for $^{18}$O + $^{70,72,74}${Ge} reactions lies between 0.212-0.214, providing a significant enhancement in the fusion cross-section compared to 1D-BPM. On the other hand, the fusion cross-section calculated for $^{18}$O + $^{148,150}$Nd, $^{18}$O + $^{150,152,154}$Sm, and $^{18}$O + $^{182,184,186}$W reactions with $2^+$ rotational state shows a noticeable increment in comparison to the 1D-BPM even at sub-barrier energies as shown in Fig. \ref{fig. 3}. The enhancement observed in the case of $^{18}$O + $^{150}$Sm, $^{18}$O + $^{148}$Nd reactions is relatively lesser than the other systems with rotational target nuclei. This is due to the smaller values of $\beta_2$ lies in between 0.164 to 0.181. The calculated results are further compared with the experimental data, wherever available. The fusion cross-section obtained using microscopic R3Y NN interaction potential is large compared to the one obtained using the WS potential. It is concluded from the results that the fusion cross-section obtained after the inclusion of vibrational and/or rotational degrees of freedom matches well with the experimental data at the above barrier energies. However, hindrance is still observed at energies below the Coulomb barrier.
\begin{figure*}
\begin{center}
\includegraphics[width=130mm,height=80mm,scale=1.5]{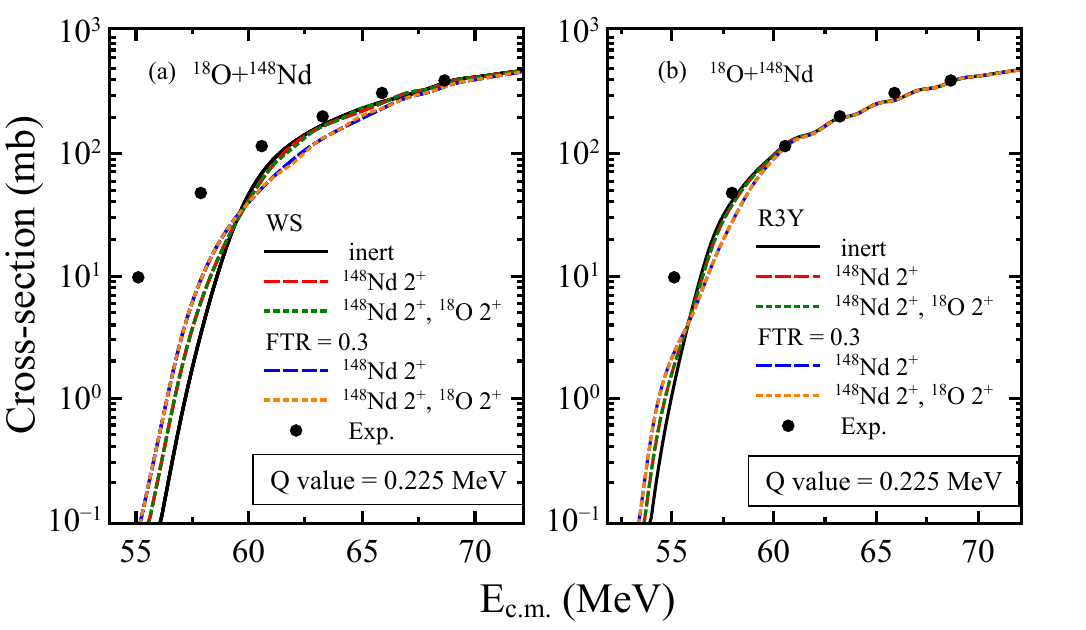}
\vspace{-0.3cm}
\caption{\label{fig. 4} (Color online) Fusion excitation functions for $^{18}$O + $^{148}$Nd reactions with Woods-Saxon potential as well as with R3Y NN potential using sharp cut-off model as shown in panel (a) and (b). The inclusion of $2n$-neutron transfer with $F_{tr}$ 0.3 fm is employed in the CCFULL code. The experimental data is taken from \cite{brod75}. }
\end{center}
\end{figure*}

\subsubsection{Neutron transfer effect on the Fusion Cross-Section}
The neutron transfer in a fusion reaction can further reduce the remaining fusion hindrance at sub-barrier energies. The code CCFULL offers a schematic representation of the pair-transfer channel by incorporating a macroscopic form factor for the coupling strength. The $Q$-values are calculated from the nuclear masses given on the NNDC website and the coupling strength ($F_{tr}$) associated with the neutron pair-transfer channel is treated as a free parameter. This parameter is optimized to achieve the best possible agreement between the theoretical and experimental fusion excitation functions, ensuring a comprehensive understanding of the system under investigation. All $^{18}$O-induced reactions discussed here have negative $Q_{1n}$ transfer and positive $Q_{2n}$ transfer, as mentioned in Table \ref{table1}. The influence of $2n$ transfer $Q$-value on the fusion cross sections caused by $^{18}$O has been examined in several studies within different mass regions \cite{beck80,brog83,brogl83,henn87,stel90,zagr03,jia16,kohl13,timm98,stef06,zhan10,kola12,jaco86,leig95} and Refs. therein using phenomenological WS potential.

However, in the present study, $Q_{2n}$ transfer is incorporated into the coupling scheme of the $^{18}$O + $^{58,60,62,64}$Ni, $^{70,72,74,76}$Ge, $^{144,148,150}$Nd, $^{144,148,150,152,154}$Sm, and $^{182,184,186}$W reactions using the microscopic R3Y NN interaction potential to observe the effect of transfer channel on the below barrier fusion cross-sections. As evident from the literature as well that internal structure effects such as nuclear shape, rotation, vibration, neck formation, and neutron transfer channel, associated with colliding systems significantly influence the fusion cross-sections at near and sub-barrier energies. Also, in our previous calculations, the microscopic nuclear potential was limited to the spherical coordinate system for calculating the fusion cross-section. Therefore, incorporating nuclear shape and rotational degrees of freedom within the microscopic RMF approach in the CCFULL code, further reduces the fusion hindrance at below-barrier energies, introducing new physics and insights in the present study. The macroscopic form factor used in the CCFULL program is mentioned in Eq. \ref{ftr}, where $F_{tr}$ is the coupling strength. In the present work, a macroscopic form factor is kept fixed at $F_{tr}$ = 0.3  in the CCFULL formalism to incorporate the pair transfer channel’s effect. This microscopic form factor provides a reasonable fit to the fusion excitation function at above barrier energy region \cite{deb20}. The dotted dashed green and purple line in Figs. \ref{fig. 2}, \ref{fig. 3} represents the cross-section with the inclusion of a $2n$ transfer channel corresponding to WS and R3Y NN interaction potential, respectively. The $2n$ transfer $Q$-values of all the considered systems are decreasing with increases with the mass number of the target nuclei, except the systems in which target nucleus is Nd. For example, in the case of $^{18}$O + $^{58,60,62,64}$Ni, and $^{18}$O + $^{70,72,74,76}$Ge reactions, the $2n$ transfer $Q$-values of $^{18}$O + $^{58}$Ni $>$ $^{18}$O + $^{60}$Ni $>$ $^{18}$O + $^{62}$Ni $>$ $^{18}$O + $^{64}$Ni, and $^{18}$O + $^{70}$Ge $>$ $^{18}$O + $^{72}$Ge $>$ $^{18}$O + $^{74}$Ge $>$ $^{18}$O + $^{76}$Ge. However, no significant enhancement is noticed in the fusion cross-section even after including the positive $Q$-value neutron transfer (PQNT) channel, except for $^{18}$O + $^{76}$Ge system as shown in Fig. \ref{fig. 2}. The inclusion of neutron transfer channel into the coupling mechanism fails to improve the match with the available experimental data. Similar trend for decrease in the $Q$-value of the neutron transfer channel with the increase in mass number of the target is followed for $^{18}$O-induced reactions having target Sm and W. The $2n$ transfer $Q$-values of $^{18}$O + $^{144}$Sm $>$ $^{18}$O + $^{148}$Sm $>$ $^{18}$O + $^{150}$Sm $>$ $^{18}$O + $^{152}$Sm $>$ $^{18}$O + $^{154}$Sm, and $^{18}$O + $^{182}$W $>$ $^{18}$O + $^{184}$W $>$ $^{18}$O + $^{186}$W. However, the value of $2n$ transfer $Q$-values in case of $^{18}$O + $^{144,148,150}$Nd reactions is $^{18}$O + $^{144}$Nd $>$ $^{18}$O + $^{148}$Nd $<$ $^{18}$O + $^{150}$Nd. The small change in the fusion cross-section has been observed with the inclusion of neutron transfer channel as compared to the excitation state of target nuclei only as shown in Figs. \ref{fig. 2}, \ref{fig. 3}. 

The discussion above reveals that a substantial $2n$ transfer $Q$-value does not necessarily lead to a notable increase in the fusion cross-section. Even a slight alteration in the deformation ($\beta_2$, $\beta_4$) of targets post $2n$ transfer could elucidate these fusion enhancements. For instance, in the reactions of $^{18}$O + $^{58,60,62,64}$Ni, there is no significant change in fusion cross-section at sub-barrier energies despite including the $2n$ transfer channel only with respect to $1^{st}$ excitation state of the target nuclei as depicted in Fig. \ref{fig. 2}(a-d). However, a slight upsurge in fusion cross-section is observed for $^{18}$O + $^{70,72,74,76}$Ge reactions at below barrier energies solely concerning the $2^+$ excitation state, as shown in Fig. \ref{fig. 2}(e-h). This is due to the minimal deviation in $\beta_2$ and $\beta_4$ values post $2n$ transfer. A marked change in fusion cross-section is noted for $^{18}$O + $^{144,148}$Nd reactions, as illustrated in Fig. \ref{fig. 3}(a) and \ref{fig. 3}(b). However, a negligible change in fusion cross-section is observed for $^{18}$O + $^{150}$Nd reaction after incorporating the $2n$ transfer channel relative to the target nuclei's $1^{st}$ excitation state only, as shown in Fig. \ref{fig. 3}(c), which can be correlated with the increase in $\beta_2$ and/or $\beta_4$ values. Likewise, the increase in fusion cross-section for $^{18}$O + $^{144,148}$Sm reactions, as depicted in Fig. \ref{fig. 3}(d), \ref{fig. 3}(e), is solely due to the increase in $\beta_2$ value. A minor increase in cross-section is observed for $^{18}$O + $^{150}$Sm reaction, but no significant enhancement in fusion cross-section is seen for $^{18}$O + $^{152,154}$Sm reactions, as shown in Fig. \ref{fig. 3}(f), \ref{fig. 3}(g), \ref{fig. 3}(h). This is because the change in $\beta_2$ and $\beta_4$ values after $2n$ transfer for $^{18}$O + $^{150}$Sm reaction is significant. Furthermore, concerning reactions involving $^{182,184,186}$W isotopes, as demonstrated in Fig. \ref{fig. 3}(i), \ref{fig. 3}(j), \ref{fig. 3}(k), a marginal increase in fusion cross-sections is observed compared to the target nuclei's $1^{st}$ excitation state. The disparity observed in the behavior of $^{18}$O + $^{148}$Nd and $^{18}$O + $^{148}$Sm reactions can be attributed to the significant change in target nuclei deformation before and after the $2n$ transfer channel. Specifically, the alteration in $\beta_2$ values before and after $2n$ transfer for $^{18}$O + $^{148}$Nd reaction is found to be 0.062 and 0.068 for $^{18}$O + $^{148}$Sm reaction. In contrast, the deviation between $\beta_4$ values is determined to be 0.030 for $^{18}$O + $^{148}$Nd reaction and 0.012 for $^{18}$O + $^{148}$Sm reaction. Remarkably, the variation in $\beta_2$ value is relatively lower in the case of $^{18}$O + $^{148}$Nd reaction, whereas the deviation in $\beta_4$ value is more pronounced for $^{18}$O + $^{148}$Sm reaction. Consequently, the investigation of deformation values in $^{18}$O + $^{148}$Nd reaction demonstrates a significant change in the deformation parameter after neutron transfer compared to other systems. It is observed from the results that fusion cross-section obtained using microscopic nuclear potential is higher compared to the Woods-Saxon (WS) potential. The computed results with positive $Q$-values using R3Y NN interaction potential exhibit better agreement with available experimental data.
\begin{figure*}
\begin{center}
\includegraphics[width=140mm,height=60mm,scale=1.5]{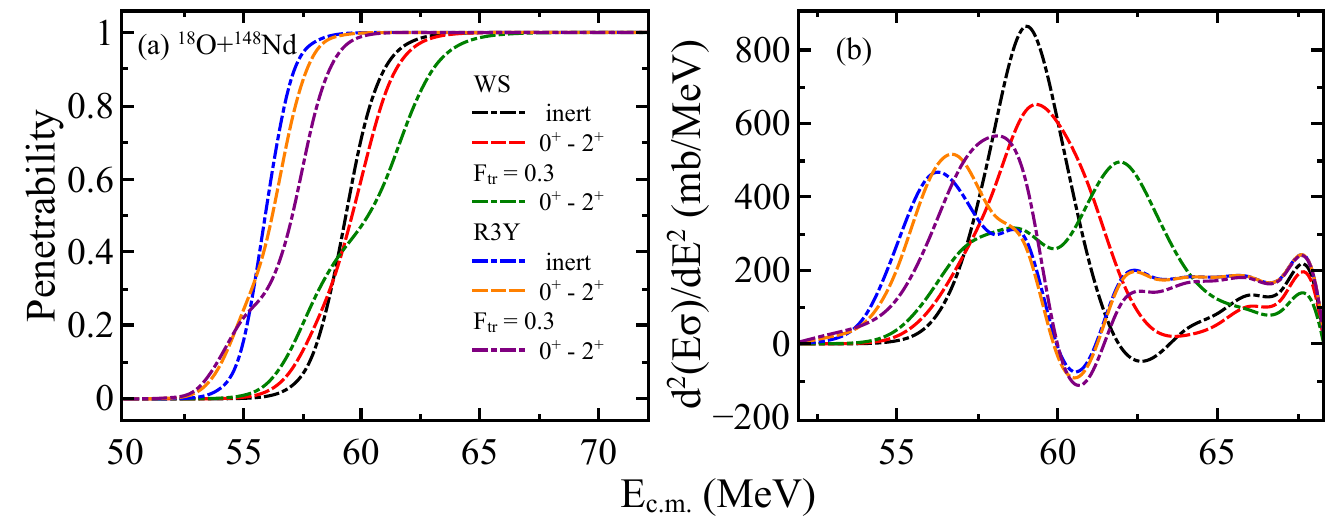}
\vspace{-0.3cm}
\caption{\label{fig. 5} (Color online) Graph for (a) penetration penetrability and (b) fusion barrier distribution as a function of $E_{c.m.}$ for $^{18}$O + $^{148}$Nd reaction.}
\end{center}
\end{figure*}

\subsubsection{Effect of Vibrational/Rotational state of interacting nuclei (Projectile/Target) on the Fusion Cross-Section}
Furthermore, we have also included the projectile’s excitation with the target nuclei to perform the CC calculations. Here, we have chosen the $^{18}$O + $^{148}$Nd reaction as an illustrative case. Similar results can be obtained for the other reactions as well. Figure \ref{fig. 4}(a) illustrates the fusion cross-section obtained using the WS potential, while Figure \ref{fig. 4}(b) displays the fusion cross-section obtained using the R3Y NN potential. The solid black line in Fig. \ref{fig. 4} represents the fusion cross-section obtained using 1D-BPM. With the inclusion of the $2^+$ vibrational state (1 phonon) of the $^{18}$O with the target’s $2^+$ excitation state (dotted green) in the calculation results in overlapping of the cross-section obtained with the $2^+$ excitation state of the target (dashed red) only. Similarly there is an overlap in the fusion cross-section obtained after the inclusion of neutron transfer channel as represented by dashed blue and dotted orange line without and with the involvement of $2^+$ vibrational state of $^{18}$O as a projectile with target nuclei. Therefore, the $2^+$ vibrational channel of $^{18}$O is not involved in the CC calculations as it is not contributing significantly. 

\subsection{Penetration Probability and Fusion Barrier Distribution}
As the nuclei fuse during the process, the nuclear matter is transferred between them in the neck region, which increases the kinetic energy of the neck constituents. This growth pattern naturally increases the fusion penetration coefficient, producing high fusion yields even when the energy is below the Coulomb barrier. Furthermore, the effect of different intrinsic degrees of freedom, like vibration and rotations, have been considered to calculate penetration probability. The relative cross-section is calculated using the coupled channel formalism. The barrier distribution is determined by the entrance flux splitting among all the coupling channels, each with a distinct barrier. To examine the effect of penetrability on the fusion barrier, we have considered the case of $^{18}$O+$^{148}$Nd reaction because the change in the fusion cross-section with shape degrees of freedom is significant $w.r.t.$ the 1D-BPM for this reaction. The change in penetration probability is further investigated as a function of $E_{c.m.}$ (MeV) to examine the effect of various incoming channels for the spherical+deformed combination choices taken into consideration, as shown in Fig. \ref{fig. 5}(a). It is observed from this Figure that the increase in states from $0^+$ to $2^+$ as well as the addition of transfer channel, penetrability substantially affects particularly at the lower barrier energies. The higher order nuclear deformation has a significant effect on the fusion barrier distributions (BDs). The fusion BD is used to examine the nature of the couplings involved in the sub-barrier fusion. The BDs were obtained through a double differentiation of the measured fusion cross-section ($\sigma_{fus}$) with respect to the incident energy, providing a comprehensive characterization of the energy dependence of fusion. Fig. \ref{fig. 5}(b) shows the fusion BD obtained for the $^{18}$O + $^{148}$Nd system. The BD have two peaks around the Coulomb barrier as compared to the 1D-BPM. This is because of the inelastic coupling of the interacting nuclei as shown in Fig. \ref{fig. 3}(b). 

It is observed from the results that on considering the $2n$ transfer channel with rotational state of the nuclei, the barrier peak shifts slightly towards higher energy than that of $2^+$ excited state of target nuclei only. Similarly, the BD's peak of inelastic state of target nuclei shifted more towards higher energy in comparison to the 1D-BPM. The incorporation of neutron transfer in the analysis results in a noticeable shift of the distribution peak towards slightly higher energies, influencing both the coupling to the 1D barrier and the coupling to collective excitations. Consequently, the presence of the positive $Q$-value neutron transfer channel (PQNT) effect becomes apparent in the present system, enhancing our understanding of fusion cross-section ($\sigma_{fus}$) interpretation. The inclusion of intrinsic degrees of freedom gives higher penetrability at lower $E_{c.m.}$ and starts decreasing with the increases in $E_{c.m.}$. \\ \\
To comprehensively assess the effect of PQNT channels on fusion enhancement in $^{18}$O-induced reactions, we thoroughly examined the available literature on $^{18}$O-induced reactions. Most of the reactions in Table \ref{table1} do not exhibit any significant enhancement attributable to the PQNT channel. However, a significant enhancement was observed in the case of $^{18}$O + $^{148}$Nd reactions due to the presence of PQNT channel. Specifically, following the $2n$ transfer process, the deformation of the target nucleus increased. This increase in deformation subsequently reduced the barrier height, thereby resulting in enhanced fusion cross-sections below the barrier. This collective evidence supports the conclusion that in systems featuring a PQNT channel, an increase in target deformation after neutron transfer corresponds to fusion enhancement. Interestingly, it is essential to note that the $Z_pZ_t$ values (product of the atomic numbers of the projectile and target) remain constant after the $2n$ transfer. Therefore, the observed increase in target deformation after transfer could be attributed to the decrease in barrier height, promoting fusion enhancement. Notably, the highlighted systems in Table \ref{table1}, exhibiting PQNT-induced enhancement, further reinforce the notion that enhancement is specifically observed when the target's deformation increases after the $2n$ transfer event. Furthermore, it should be emphasized that a higher $Q$-value for the $2n$ transfer does not guarantee a more substantial enhancement. The increase in the fusion cross-section depends upon the variation of deformation parameter after $2n$ transfer channel. Specifically, the resulting fusion cross-section remains unchanged when the deformation parameter decreases after the neutron channel. On the other hand, the fusion cross-section for the system under study shows noticeable enhancement if the deformation parameter increases after $2n$ transfer channel. Thus, when analyzing these reactions, it is crucial to consider the relationship between the PQNT channel, after transfer deformation changes, and resulting fusion enhancement.

\section{Summary and Conclusions}
\label{result} \noindent
This study focuses on the enhanced fusion cross-sections observed in heavy-ion-induced reactions, particularly at energies below the barrier. The one-dimensional barrier penetration model with the bare nuclear potential is insufficient to explain this enhancement, suggesting the involvement of factors like deformation and coupling to low-lying inelastic states. To gain a better understanding, the present work investigates below-barrier fusion enhancement in systems with positive $Q$-values for neutron transfer. The microscopic R3Y nucleon-nucleon interaction potential, obtained from the relativistic mean-field (RMF) approach within the NL3$^*$ parameter set, is employed in the present analysis. The measured fusion cross-sections for $^{18}$O-induced reactions have been thoroughly examined using the coupled channel (CCFULL) approach. Our study focuses on reactions involving  $^{18}$O + $^{58,60,64}${Ni}, $^{18}$O + $^{74}${Ge}, $^{18}$O + $^{148}${Nd}, $^{18}$O + $^{150}${Sm}, and $^{18}$O + $^{182,184,186}$W, for which experimental data is available. The analysis considers the influence of couplings on collective excitations, including the target's vibrational and/or rotational states and the projectile's vibrational $2^+$ state. To successfully reproduce the data, we incorporated a two-neutron $(2n)$ transfer channel along with these couplings in the calculation. Furthermore, a comparison is made between the results obtained using microscopic R3Y NN interaction potential and the Woods-Saxon potential. The relativistic R3Y NN potential is noted to yield lower barrier heights compared to the WS potential for all considered reaction systems. Notably, even a minor alteration in the height of the R3Y potential barrier exerts a significant effect on the cross-section, resulting in a significant increase in the cross-section below the Coulomb barrier. Our results indicate that the R3Y NN potential, which considers the effects of meson coupling, demonstrates better agreement with experimental fusion cross-sections than the WS potential. Specifically, three reactions, namely, $^{18}$O + $^{150}${Nd}, and $^{18}$O + $^{152,154}${Sm}, exhibit large cross-sections in the below-energy region with the inclusion of the neutron transfer channel compared to the 1D-BPM. Encouraged by these results, we extend our work to predict the outcome of previously unexplored $^{18}$O-induced reactions, including $^{18}$O + $^{62}${Ni}, $^{18}$O + $^{70,72,76}${Ge}, $^{18}$O + $^{144,150}${Nd}, $^{18}$O + $^{144,148,152,154}${Sm}. Interestingly, using the R3Y NN potential, our studies demonstrate the critical role of positive $Q$-value neutron transfer in enhancing the sub-barrier fusion cross-sections, especially in the $^{18}$O + $^{148}${Nd} reaction. Additionally, systematic studies on $^{18}$O-induced reactions in different mass regions suggest that PQNT reactions have higher cross-sections due to increased deformation from neutron transfer. However, further experimental investigations on transfer reactions are required to gain a more comprehensive understanding of the effects of positive $Q$-value neutron transfer.

\section*{Acknowledgments}
\noindent
This work has been supported by the Science Engineering Research Board (SERB) File No. CRG/2021/001229, Sao Paulo Research Foundation (FAPESP) Grant 2017/05660-0, FOSTECT Project No. FOSTECT.2019B.04, and the Hungarian National Research, Development and Innovation Office (NKFIH K134197). The authors acknowledge Prof. A. Vitturi for
useful discussions.

\end{document}